%% file: Manuscript.tex
\documentclass[twocolumn]{aastex631}

\input{Commands}

\input{Packages}

\shorttitle{Optical and radio counterpart of \thisEP}
\shortauthors{Gillanders, Rhodes \& Srivastav et al.}

\begin{document}

\title{Discovery of the optical and radio counterpart to the fast X-ray transient \thisEP}

\correspondingauthor{
    \\ James~H.~Gillanders (\href{mailto: james.gillanders@physics.ox.ac.uk}{james.gillanders@physics.ox.ac.uk}),
    \\ Lauren Rhodes (\href{mailto: lauren.rhodes@physics.ox.ac.uk}{lauren.rhodes@physics.ox.ac.uk}) \&
    \\ Shubham Srivastav (\href{mailto: shubham.srivastav@physics.ox.ac.uk}{shubham.srivastav@physics.ox.ac.uk}).
}

\author[0000-0002-8094-6108]{J.~H.~Gillanders}
\thanks{These authors contributed equally.}
\affil{Astrophysics sub-Department, Department of Physics, University of Oxford, Keble Road, Oxford, OX1 3RH, UK}
\author[0000-0003-2705-4941]{L.~Rhodes}
\thanks{These authors contributed equally.}
\affil{Astrophysics sub-Department, Department of Physics, University of Oxford, Keble Road, Oxford, OX1 3RH, UK}
\author[0000-0003-4524-6883]{S.~Srivastav}
\thanks{These authors contributed equally.}
\affil{Astrophysics sub-Department, Department of Physics, University of Oxford, Keble Road, Oxford, OX1 3RH, UK}
\author[0000-0002-0426-3276]{F.~Carotenuto}
\affil{Astrophysics sub-Department, Department of Physics, University of Oxford, Keble Road, Oxford, OX1 3RH, UK}
\author[0000-0002-7735-5796]{J.~Bright}
\affil{Astrophysics sub-Department, Department of Physics, University of Oxford, Keble Road, Oxford, OX1 3RH, UK}
\author[0000-0003-1059-9603]{M.~E.~Huber}
\affiliation{Institute for Astronomy, University of Hawai'i, 2680 Woodlawn Drive, Honolulu, HI 96822, USA}
\author[0000-0002-0504-4323]{H.~F.~Stevance}
\affil{Astrophysics sub-Department, Department of Physics, University of Oxford, Keble Road, Oxford, OX1 3RH, UK}
\affil{Astrophysics Research Centre, School of Mathematics and Physics, Queen's University Belfast, BT7 1NN, UK}
\author[0000-0002-8229-1731]{S.~J.~Smartt}
\affil{Astrophysics sub-Department, Department of Physics, University of Oxford, Keble Road, Oxford, OX1 3RH, UK}
\affil{Astrophysics Research Centre, School of Mathematics and Physics, Queen's University Belfast, BT7 1NN, UK}
\author{K.~C.~Chambers}
\affiliation{Institute for Astronomy, University of Hawai'i, 2680 Woodlawn Drive, Honolulu, HI 96822, USA}
\author[0000-0002-1066-6098]{T.-W.~Chen}
\affil{Graduate Institute of Astronomy, National Central University, 300 Jhongda Road, 32001 Jhongli, Taiwan}
\author{R.~Fender}
\affil{Astrophysics sub-Department, Department of Physics, University of Oxford, Keble Road, Oxford, OX1 3RH, UK}
\affil{Department of Astronomy, University of Cape Town, Private Bag X3, Rondebosch 7701, South Africa}
\author[0000-0003-2734-1895]{A.~Andersson}
\affil{Astrophysics sub-Department, Department of Physics, University of Oxford, Keble Road, Oxford, OX1 3RH, UK}
\author[0000-0002-4033-3139]{A.~J.~Cooper}
\affil{Astrophysics sub-Department, Department of Physics, University of Oxford, Keble Road, Oxford, OX1 3RH, UK}
\author[0000-0001-5679-0695]{P.~G.~Jonker}
\affil{Department of Astrophysics/IMAPP, Radboud University, P.O.~Box 9010, 6500 GL, Nijmegen, The Netherlands}
\author[0009-0009-0079-2419]{F.~J.~Cowie}
\affil{Astrophysics sub-Department, Department of Physics, University of Oxford, Keble Road, Oxford, OX1 3RH, UK}
%
%
\author{T.~de~Boer}
\affil{Institute for Astronomy, University of Hawai'i, 2680 Woodlawn Drive, Honolulu, HI 96822, USA}
\author[0000-0002-9986-3898]{N.~Erasmus}
\affil{South African Astronomical Observatory, PO Box 9, Observatory 7935, Cape Town, South Africa} 
\author[0000-0003-1916-0664]{M.~D.~Fulton}
\affil{Astrophysics Research Centre, School of Mathematics and Physics, Queen's University Belfast, BT7 1NN, UK}
\author[0000-0003-1015-5367]{H.~Gao}
\affil{Institute for Astronomy, University of Hawai'i, 2680 Woodlawn Drive, Honolulu, HI 96822, USA}
\author{J.~Herman}
\affil{Institute for Astronomy, University of Hawai'i, 2680 Woodlawn Drive, Honolulu, HI 96822, USA}
\author[0000-0002-7272-5129]{C.-C.~Lin}
\affil{Institute for Astronomy, University of Hawai'i, 2680 Woodlawn Drive, Honolulu, HI 96822, USA}
\author{T.~Lowe}
\affil{Institute for Astronomy, University of Hawai'i, 2680 Woodlawn Drive, Honolulu, HI 96822, USA}
\author[0000-0002-7965-2815]{E.~A.~Magnier}
\affil{Institute for Astronomy, University of Hawai'i, 2680 Woodlawn Drive, Honolulu, HI 96822, USA}
\author[0000-0003-2736-5977]{H.-Y.~Miao}
\affil{Graduate Institute of Astronomy, National Central University, 300 Jhongda Road, 32001 Jhongli, Taiwan}
\author{P.~Minguez}
\affil{Institute for Astronomy, University of Hawai'i, 2680 Woodlawn Drive, Honolulu, HI 96822, USA}
\author[0000-0001-8385-3727]{T.~Moore}
\affil{Astrophysics Research Centre, School of Mathematics and Physics, Queen's University Belfast, BT7 1NN, UK}
\author[0000-0001-8771-7554]{C.-C.~Ngeow}
\affil{Graduate Institute of Astronomy, National Central University, 300 Jhongda Road, 32001 Jhongli, Taiwan}
\author[0000-0002-2555-3192]{M.~Nicholl}
\affil{Astrophysics Research Centre, School of Mathematics and Physics, Queen's University Belfast, BT7 1NN, UK}
\author[0000-0001-8415-6720]{Y.-C.~Pan}
\affil{Graduate Institute of Astronomy, National Central University, 300 Jhongda Road, 32001 Jhongli, Taiwan}
\author{G.~Pignata}
\affil{Instituto de Alta Investigación, Universidad de Tarapacá, Arica, Casilla 7D, Chile}
\author{A.~Rest}
\affil{Space Telescope Science Institute, 3700 San Martin Drive, Baltimore, MD 21218, USA}
\affil{Department of Physics and Astronomy, Johns Hopkins University, Baltimore, MD 21218, USA}
\author[0000-0002-6527-1368]{X.~Sheng}
\affil{Astrophysics Research Centre, School of Mathematics and Physics, Queen's University Belfast, BT7 1NN, UK}
\author[0000-0001-8605-5608]{I. A. Smith}
\affil{Institute for Astronomy, University of Hawai’i, 34 Ohia Ku St., Pukalani, HI 96768-8288, USA}
\author[0000-0001-9535-3199]{K.~W.~Smith}
\affil{Astrophysics Research Centre, School of Mathematics and Physics, Queen's University Belfast, BT7 1NN, UK}
\author{J.~L.~Tonry}
\affil{Institute for Astronomy, University of Hawai'i, 2680 Woodlawn Drive, Honolulu, HI 96822, USA}
\author[0000-0002-1341-0952]{R.~J.~Wainscoat}
\affil{Institute for Astronomy, University of Hawai'i, 2680 Woodlawn Drive, Honolulu, HI 96822, USA}
\author{J.~Weston}
\affil{Astrophysics Research Centre, School of Mathematics and Physics, Queen's University Belfast, BT7 1NN, UK}
\author{S.~Yang}
\affil{Henan Academy of Sciences, Zhengzhou 450046, Henan, China}
\author[0000-0002-1229-2499]{D.~R.~Young}
\affil{Astrophysics Research Centre, School of Mathematics and Physics, Queen's University Belfast, BT7 1NN, UK}

\begin{abstract}
Fast X-ray Transients (FXTs) are extragalactic bursts of soft X-rays first identified $\gtrsim 10$~years ago. Since then, nearly 40 events have been discovered, although almost all of these have been recovered from archival {\it Chandra} and {\it XMM-Newton} data. To date, optical sky surveys and follow-up searches have not revealed any multi-wavelength counterparts. The Einstein Probe, launched in January 2024, has started surveying the sky in the soft X-ray regime ($0.5 - 4$\,keV) and will rapidly increase the sample of FXTs discovered in real time. Here, we report the first discovery of both an optical and radio counterpart to a distant FXT, the fourth source publicly released by the Einstein Probe. We discovered a fast-fading optical transient within the 3~arcmin localisation radius of \thisEP\ with the all-sky optical survey ATLAS, and our follow-up Gemini spectrum provides a redshift, $z = 4.859 \pm 0.002$. Furthermore, we uncovered a radio counterpart in the S-band (3.0\,GHz) with the MeerKAT radio interferometer. The optical (rest-frame UV) and radio luminosities indicate the FXT most likely originates from either a long gamma-ray burst or a relativistic tidal disruption event. This may be a fortuitous early mission detection by the Einstein Probe or may signpost a mode of discovery for high-redshift, high-energy transients through soft X-ray surveys, combined with locating multi-wavelength counterparts. 
\end{abstract}


\section{Introduction} \label{sec:Introduction}

In the last decade, a few tens of fast X-ray transients have been discovered with \textit{Chandra}, \textit{XMM-Newton} and \textit{eROSITA} \citep[see \eg,][]{jonker13, glennie15, bauer17, alp_larssson_20, Quirola2022, Quirola2023}. These bursts are soft ($0.3 - 10$\,keV) and exhibit a wide range of timescales, lasting from $\sim 10^1 - 10^4$~seconds, with a variety of astrophysical interpretations having been invoked to explain their properties. 

Events such as CDF-S\,XT2 \citep{xue19}, XRT\,210423 \citep{ai_zhang_21, Eappachen2023a} and CDF-S\,XT1 \citep{sarin21} have been interpreted as resulting from a binary neutron star (BNS) merger. CDF-S\,XT2 and XRT\,210423 both showed a clear plateau in the X-ray lightcurve, followed by a sharp drop, consistent with model predictions for a rapidly spinning magnetar remnant. On the other hand, XRT\,000519 showed precursor X-ray emission 4000 and 8000\,s before the main flare \citep{jonker13}, the timescale of which agrees with the expected orbital timescale of a white dwarf (WD) spiralling towards an intermediate-mass black hole (IMBH) on an eccentric orbit \citep{macleod16}. \cite{glennie15} found two FXTs in archival \textit{Chandra} data, and reported an infrared (IR) Galactic counterpart at a distance of 80\,pc for one of them (XRT\,120830). They interpret this FXT to be consistent with an M-dwarf super flare, but the other had no detected counterpart. 

\cite{alp_larssson_20} reported 12 FXTs from \textit{XMM-Newton} and from inference of potential hosts they interpret the FXTs as emission from shock breakout in Wolf-Rayet stars within a dense circumstellar medium or (favoured in two cases) red supergiant progenitors. \cite{eappachen24} showed that seven of these have plausible host galaxies with spectroscopic redshifts $0.098< z < 0.645$, with one being a likely Galactic flare star. They proposed one FXT (XRT\,110621) is consistent with being a supernova shock breakout (SBO) but the spectroscopic redshifts of the others showed that their peak X-ray luminosities were above that deemed feasible for supernova SBOs. \cite{Soderberg2008} report an X-ray detection which they associate with the SBO from \SNxx{2008D}, at a distance of 27\,Mpc \citep[see also][]{Chevalier2008, Mazzali2008, Modjaz2009}. \cite{eappachen24} also searched for contemporaneous optical counterparts in the Pan-STARRS and ATLAS wide-field surveys but found none. The detection limits range in depth (from $m_w \simeq 22$ to $m_o \simeq 18.4$; AB mags), and the delay between the bursts and the observations range from $1 - 170$~days. The most stringent limit on any contemporaneous optical emission remains the serendipitous observation of the location of CDF-S\,XT\,1 with the Very Large Telescope (VLT) just 80~minutes after the burst \citep{bauer17}. With this observation, no associated optical counterpart -- or host galaxy -- was detected, down to a limiting $R$-band magnitude, $m_R > 25.7$~AB~mag.

The discovery and rapid follow-up of FXTs is expected to accelerate since the launch of the Einstein Probe \citep[EP;][]{Yuan2022_EinsteinProbe} on January~9~2024. With its instantaneous wide field of view of 3600~square degrees, the mission is designed to survey the available night-time sky several times per day in the soft X-ray regime ($0.5 - 4$\,keV), and to follow-up detected transients. During its commissioning phase, it has already proven to be a valuable discovery instrument, with four new X-ray transient sources reported by mid-March. The first, EPW\,20240219aa \citep{ATel16463}, has had no multi-wavelength counterpart identified, but an association with a sub-threshold Fermi Gamma-ray Burst Monitor (GBM) detection has been made \citep{2024GCN.35773....1Z, 2024GCN.35776....1F}, suggesting it may be a GRB event. The following two EP transients released are almost certainly Galactic. EPW\,20240305aa \citep{ATel16509} has been well localised by \textit{Swift}/XRT \citep{ATel16514}, and is coincident with a Gaia DR3 star \citep[late A-type or early F-type;][]{ATel16529}, with radio emission observed by ATCA \citep{ATel16555}. EP\,240309a was detected as a highly variable X-ray source \citep[previously detected by \textit{XMM-Newton}, \textit{Swift} and \textit{eROSITA};][]{ATel16546}, and has been confirmed as a cataclysmic variable with an orbital period of 3.76\,hr \citep{ATel16549, ATel16554}. 

The fourth bright transient source publicly released by the Einstein Probe mission, \thisEP, was detected on 2024 March 15 20:10:44 UTC ($T_0 = $~MJD~60384.84079) by the wide-field X-ray telescope \citep{Zhang2024_GCN35931}. The EP team reported that the event lasted 1600 seconds, with a peak flux, $f_{\rm X} \sim 3 \times 10^{-9}$\,\ergscm\ in the $0.5 - 4$\,keV band. No previously known X-ray sources were identified in the 3~arcmin localisation radius, making it a candidate extragalactic FXT.

In this Letter, we report the discovery of the optical and radio transient associated with \thisEP, the first time multi-wavelength counterparts of a `distant' ($D \gtrsim 100$\,Mpc) extragalactic FXT have been recorded. Throughout this paper we assume $\Lambda$CDM cosmology with a Hubble constant, $H_0 = 67.7$\,km\,s$^{-1}$\,Mpc$^{-1}$, $\Omega_{\rm M} = 0.309$ and \mbox{$\Omega_{\Lambda} = 0.691$} \citep{2016A&A...594A..13P}. We also assume a line-of-sight Milky Way extinction of \mbox{$E(B-V) = 0.042$} AB mag, which corresponds to \mbox{$A_V = 0.130$} AB mag \citep{Schlafly2011}.

\begin{figure*}
    \centering
    \includegraphics[width=0.9\linewidth]{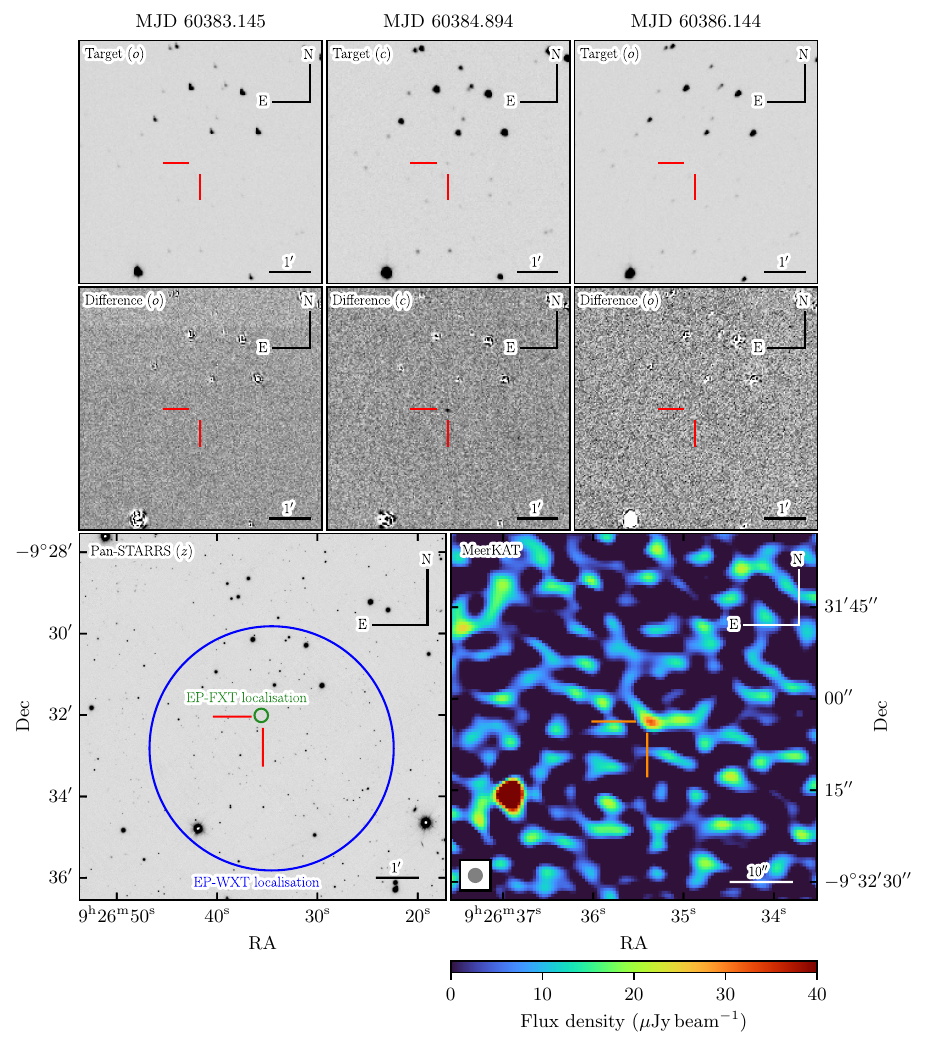}
    \caption{
        The location of the counterparts to \thisEP, marked with cross-hairs (red for the optical location from ATLAS, orange for the radio location from MeerKAT).
        \textit{Top panels:} ATLAS stacked ($4 \times 30$\,s) target images of the field of \thisAT. Left to right: $o$-band observation taken at $T_0 - 1.696$\,d, detection $c$-band observation taken at $T_0 + 0.054$\,d, and the subsequent $o$-band observation, taken at $T_0 + 1.304$\,d. \thisAT\ faded below the detection threshold of ATLAS in $\lesssim 1.2$~days ($m_o > 21.1$ AB mag; $2 \sigma$ upper limit).
        \textit{Middle panels:} Same as the top panels, but here we present the difference images. In the centre panel, the presence of \thisAT\ is unmistakable.
        \textit{Bottom left panel:} Pan-STARRS $z$-band image of the field of \thisAT, with the localisation regions from the detections of \thisEP\ by the EP-WXT \citep[blue; from][]{Zhang2024_GCN35931} and EP-FXT \citep[green; from][]{Chen2024_GCN35951} overlaid, to illustrate the spatial coincidence with the optical counterpart \thisAT.
        \textit{Bottom right panel:} MeerKAT radio image of the field of \thisEP. There is clear evidence for a bright radio source in the image, coincident with \thisEP. The beam size for the observations is $3.5'' \times 3.5''$, illustrated by the stamp in the lower left corner. Note the much smaller scale in this image compared with the optical images.
    }
    \label{fig:AT2024eju detection images}
\end{figure*}

\section{Multi-wavelength counterpart discovery and follow-up} \label{sec:Discovery and follow-up}

\subsection{Discovery of the optical counterpart with ATLAS} \label{sec:Discovery and follow-up - ATLAS}

The Asteroid Terrestrial-impact Last Alert System \citep[ATLAS;][]{Tonry2018_ATLAS} is a quadruple 0.5-m telescope system, operating a wide-field all-sky survey. ATLAS continually surveys the sky, typically four times in 24\,hr when all four units are operating normally, and we promptly process the data to search for extragalactic transients \citep{Smith2020_ATLAS}. During its normal survey operations, ATLAS observed the localisation region of \thisEP\ at MJD 60384.894,\footnote{Here (and for all other optical imaging observations), we quote the epoch of observation as the midpoint of the exposure.} corresponding to \mbox{$T_0 + 1.28$\,hr} (note that the first of the four 30\,s exposures was obtained at MJD~60384.88673, or $T_0 + 1.10$\,hr). Recall $T_0$ is the time of the detection from the Einstein Probe \citep[MJD 60384.84079;][]{Zhang2024_GCN35931}.

Observations were performed by the Sutherland unit in South Africa, with $4 \times 30$\,s exposures obtained using the \textit{cyan}, or $c$, filter (analogous to the Pan-STARRS/SDSS $g + r$ filters). During automated image processing \citep[outlined by][]{Smith2020_ATLAS} the observations were reduced and calibrated photometrically and astrometrically with the reference catalogue RefCat2 \citep{Tonry2018_REFCAT}, and a reference image was subtracted. We registered the optical transient \thisAT\ (ATLAS24dsx) with sky-coordinates of RA~$= +141.64763$, Dec~$= -9.53401$ ($9^{\rm h} 26^{\rm m} 35.43^{\rm s}$, $-9^\circ 32' 02.4''$), and an observed magnitude, \mbox{$m_c = 19.38 \pm 0.08$~AB~mag} on the Transient Name Server \citep{TNSTR2024eju}. With no detection of the source in ATLAS images 1.75\,d before, no historical variability, and a 0.8~arcmin spatial separation, we reported this as a plausible counterpart to \thisEP\ \citep{Srivastav2024_GCN35932}.

In Figure~\ref{fig:AT2024eju detection images}, we present the nightly stacked ($4 \times 30$\,s) target and difference images from ATLAS for the detection epoch, and the neighbouring epochs immediately pre- and post-detection. The presence of the transient on MJD~60384.894 is unmistakable, with no evidence for \thisAT\ in the most recent previous observation (indicating no pre-existing transient activity), and no evidence in the subsequent observation (indicating its rapid fade). Figure~\ref{fig:AT2024eju detection images} visually highlights how rapidly \thisAT\ rose and subsequently faded.

\subsection{Optical photometric follow-up} \label{sec:Discovery and follow-up - Photometry}

After the initial discovery with ATLAS, we triggered rapid multi-band follow-up imaging observations with the Pan-STARRS telescopes, the Liverpool Telescope and the Lulin Observatory. All three observatories were triggered and on-source within $24 - 36$~hours. 

We used the 40-cm SLT located at Lulin Observatory, Taiwan, to obtain $r$-band images of the field of \thisEP\ as part of the Kinder project \citep{Chen2021_KINDER}. The initial observation with SLT began at MJD~60385.673, or $T_0 + 0.832$\,d. We successfully recovered \thisAT\ in the images, albeit with a marginal detection \citep{Chen2024_GCN35938}, indicating a fast fade within the first 24\,hr of the FXT discovery. Subsequently, we conducted continuous observations of \thisAT\ using both SLT and the Lulin One-meter Telescope (LOT) with $i$-band imaging. We employed the Kinder pipeline \citep{kinderpip} to conduct PSF photometry for \thisAT\ without template subtraction. The derived magnitudes and $2 \sigma$ upper limits were determined by calibrating against Pan-STARRS1 field stars in the AB system. 

The 2-m Liverpool Telescope \citep[LT;][]{Steele2004_LT} was triggered under the program PL24A28 (PI: S.~Srivastav). Images were obtained in $gri$-bands commencing on MJD 60385.848, corresponding to \mbox{$T_0 + 1.007$\,d}. While the observing conditions were poor and the optical counterpart was not detected, our upper limit from non-detections confirmed the rapidly fading nature of \thisAT\ \citep[see][]{Srivastav2024_GCN35933}. Another set of $iz$-band images were obtained the next night, at MJD~60386.946, in better conditions. However, given the continuing rapid fade, \thisAT\ was only detected in $i$-band. The derived magnitudes and $2 \sigma$ upper limits from the LT images were estimated using the python-based \texttt{Photometry Sans Frustration} (\texttt{PSF}) code\footnote{\url{https://github.com/mnicholl/photometry-sans-frustration}} \citep{Nicholl2023_PSF}. 
 
Pan-STARRS observations commenced on MJD~60386.324, or \mbox{$T_0 + 1.483$~days}. The Pan-STARRS (PS) system is a twin 1.8-m telescope system (Pan-STARRS1 and Pan-STARRS2), both situated atop Haleakala mountain on the Hawaiian island of Maui \citep{Chambers2016arXiv_PanSTARRS1}. All observations of \thisAT\ were performed with Pan-STARRS1 (PS1), which has a 1.4~gigapixel camera and  0.26~arcsec pixels. This provides a focal plane with a diameter of 3.0~degrees, and a field-of-view area of 7.06~square degrees, which can be imaged with the \grizy\ filter system \citep[as described by][]{Tonry2012}. Images were processed with the Image Processing Pipeline \citep[IPP;][]{magnier2020a, waters2020}. The individual exposure frames were astrometrically and photometrically calibrated \citep{magnier2020c} and overlapping exposures co-added together with median clipping applied (to produce stacks) on which PSF photometry was performed \citep{magnier2020b}. We commenced targeted observations on MJD~60386.324 (\mbox{$T_0 + 1.483$\,d}). Two epochs of observations were obtained on the first night, with the initial \grizy\ followed $\sim 1.4$\,hr later by \izy\ imaging. We dropped the $gr$-bands from all subsequent follow-up due to the non-detections in our first epoch.

Finally, we obtained an epoch of late-time $iz$-band imaging with the Gemini-North/GMOS-N instrument, under the program ID GN-2024A-Q-221 (PI:~M.~Huber), at MJD 60403.257 ($T_0 + 18.42$\,d). These observations were reduced using the \texttt{DRAGONS} pipeline \citep{Labrie2023_DRAGONS, DRAGONS_zenodo}, and following standard recipes. \thisAT\ was not detected in these deep stacked images, and the $2 \sigma$ upper limits were again derived using the \texttt{PSF} code.

\begin{figure*}
    \centering
    \includegraphics[width=\linewidth]{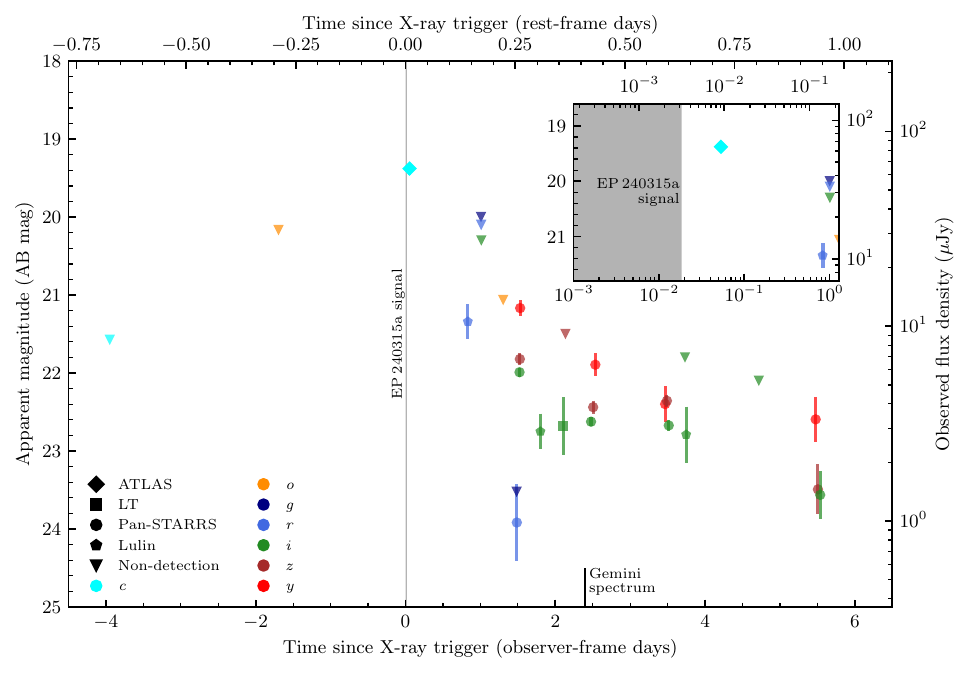}
    \caption{
        Optical photometry of \thisAT. All non-detections are represented by downward-pointing triangles (and correspond to $2 \sigma$ upper limits). Error bars correspond to $1 \sigma$ values. We present the data both in observer and rest frame, and the flux in both magnitude and $f_\nu$ space. The grey band  represents the duration of the initial X-ray detection reported by the Einstein Probe \citep[$\approx 1600$\,s; see][]{Zhang2024_GCN35931}. The epoch of spectroscopic observation with Gemini-North has also been marked (vertical black line).
        \textit{Inset panel:} A zoom-in of the first $\sim 1.5$~days to emphasise how close our ATLAS $c$-band detection was to the initial X-ray detection ($\lesssim 80$~observer-frame minutes, $\lesssim 800$~rest-frame seconds). Note we do not plot the late-time ($T_0 + 18.42$\,d, observer frame) Gemini non-detections.
    }
    \label{fig:AT2024eju optical lightcurve}
\end{figure*}

The full optical lightcurve information, including our ATLAS, Lulin, LT, Pan-STARRS and Gemini photometry, is presented in Figure~\ref{fig:AT2024eju optical lightcurve} and Table~\ref{tab:Photometry}.

\subsection{Spectroscopic observation with Gemini and redshift measurement} \label{sec:Discovery and follow-up - Spectra}

In addition to the rapidly acquired photometric data,  we obtained a spectrum of the optical counterpart to \thisEP, commencing on MJD~60387.236 \mbox{($T_0+2.395$~days)}. Our observation was carried out using the Gemini-North/GMOS-N instrument under program ID \mbox{GN-2024A-Q-128} (PI:~M.~Huber) using the R400 grating and $1''$ slit width, which provided coverage over the \mbox{$\approx 4200 - 9100$\,\AA} wavelength range. Our observation was split into a number of sub-exposures, with a total on-target exposure time of 8880\,s.

\begin{figure*}
    \centering
    \includegraphics[width=\linewidth]{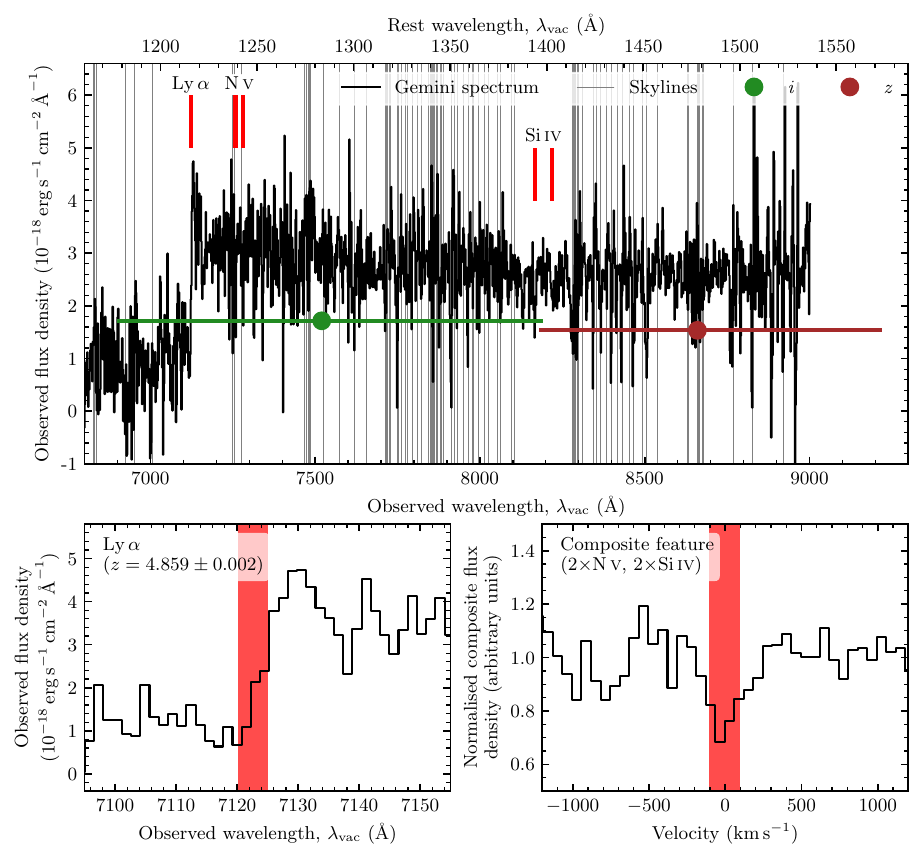}
    \caption{
        \textit{Upper panel:} Gemini-North/GMOS-N telluric-corrected spectrum of \thisEP/\thisAT. 
        The Pan-STARRS $iz$-band observations taken $\approx0.1$~days after the spectral observations are overlaid. The absorption lines from which we estimate the redshift of the system have been marked (vertical red lines). Prominent skylines \citep[from][]{Hanuschik2003} have been marked with vertical grey lines.
        \textit{Lower left panel:} A zoom-in on the region of strong \mbox{Lyman-$\alpha$} absorption (the width of the red band is representative of our redshift uncertainty).
        \textit{Lower right panel:} Composite spectrum of \thisAT, constructed from the profiles of the four absorption features we used to measure the redshift to \thisEP\ (\mbox{N\,\V\ $\lambda \lambda 1238.821, 1242.804$} and \mbox{Si\,\IV\ $\lambda \lambda 1393.755, 1402.770$}). Note this composite spectrum has been normalised and transformed to velocity space. The width of the red band is again representative of our estimated redshift uncertainty.
    }
    \label{fig:Gemini spectrum}
\end{figure*}

We reduced our Gemini observation using the \texttt{DRAGONS} pipeline \citep{Labrie2023_DRAGONS, DRAGONS_zenodo} and following standard recipes, with the reduced spectrum calibrated against a standard star. There are a number of narrow absorption lines evident in the spectrum that can be used to estimate the redshift to the system. The reduced spectrum is shown in the top panel of Figure~\ref{fig:Gemini spectrum}. We fit four of these lines as Gaussian absorption components, and estimate the centroids of the features. We find that the four absorption features are centred at \mbox{$\approx 7259$}, 7270, 8167 and 8222\,\AA, which we propose are produced by the N\,\V\ $\lambda \lambda 1238.821, 1242.804$ and Si\,\IV\ $\lambda \lambda 1393.755, 1402.770$ transitions, respectively. With these line identifications, we estimate the redshift of \thisAT\ to be $z = 4.859 \pm 0.002$. There is evidence for prominent Lyman-$\alpha$ absorption at $\approx 7125$\,\AA, in good agreement with our redshift estimate (see bottom left panel of Figure~\ref{fig:Gemini spectrum}). In the bottom-right panel of Figure~\ref{fig:Gemini spectrum}, we show a composite spectrum of the four absorption features from which we have estimated our redshift. Our derived redshift value is in line with measurements from two GCNs released after the discovery of \thisAT\ \citep[$z \approx 4.859$; see][]{Saccardi2024_GCN35936, Quirola2024_GCN35960}.

\subsection{Radio observations} \label{sec:Discovery and follow-up - Radio}

We observed the position of \thisEP\ with the MeerKAT radio telescope. The observation was performed as part of program SCI-20230907-JB-01 (PI:~J.~Bright). MeerKAT is a radio interferometer located in the Karoo desert in South Africa, and a precursor of the Square Kilometre Array (SKA). The instrument consists of $64 \times 13.5$-m antennas that are currently equipped with UHF, L-band and S-band receivers, covering the $0.5 - 3.5$\,GHz frequency range. Characterised by a dense core and with a longest baseline of 8\,km, the array offers an excellent snap-shot \textit{uv}-coverage, a large field of view (1.69 square degrees) and $\sim \mu$Jy sensitivity \citep{Camilo2018, Jonas2018}. We observed \thisEP\ with MeerKAT starting on MJD 60387.703 ($T_0 + 2.86$\,d), for a total on-source time of 42~minutes. We observed at a central frequency of 3.06\,GHz (S-band, S3), with a total bandwidth of 875\,MHz. PKS~J1939--6342 and 3C237 were used as flux and complex gain calibrators, respectively. The data were reduced with the \texttt{OxKAT} pipeline \citep{oxkat}, which performs standard flagging, calibration and imaging using \texttt{tricolour} \citep{Hugo_2022}, \texttt{CASA} \citep{CASA_team_2022} and \texttt{WSCLEAN} \citep{Offringa_wsclean}, respectively. Specifically, for the imaging part we adopted a Briggs weighting scheme with a $-0.3$ robust parameter, yielding a $3.5\arcsec \times 3.5\arcsec$ beam and $8 \, \mu$Jy\,beam$^{-1}$ rms noise in the target field. We clearly detected a point source at the position of the optical counterpart \thisAT\ \citep[first announced by][see also Figure~\ref{fig:AT2024eju detection images}]{Carotenuto2024_GCN35961}. Fitting for a point source in the image plane, we measure a flux density of $34 \pm 5 \, \mu$Jy\,beam$^{-1}$.

Upon the discovery of a radio counterpart with the MeerKAT radio telescope, we obtained a rapid response time request with the \textit{enhanced}-Multi-Element Radio Linked Interferometer Network (\textit{e}-MERLIN, DD17004; PI:~L.~Rhodes). \textit{e}-MERLIN is a UK-based radio interferometer with a maximum baseline of 217\,km, and seven dishes spanning $25 - 76$\,m in diameter. The facility can observe at L-, C- and K-band. Given the improved phase stability and sensitivity, we requested that our observation be made at C-band with a central frequency of 5.08\,GHz and a bandwidth of 0.51\,GHz. We obtained two observations; the first commenced on MJD~60389.736 ($T_0 + 4.90$\,d) and finished on MJD~60391.083 ($T_0 + 6.24$\,d), while the second started on MJD~60396.708 ($T_0 + 11.87$\,d) and finished on MJD~60398.042 ($T_0 + 13.20$\,d), each with a break in the middle whilst the target was below the horizon. The observations consisted of a series of six-minute scans of the target field, followed by two minutes on the phase calibrator (0933-0819). The target-phase cal loops were book-ended by visits to the flux and bandpass calibrators (J1331+3030 and J1407+2827, respectively). The data were flagged, calibrated and imaged using the \textit{e}-MERLIN pipeline\footnote{\url{https://github.com/e-merlin/eMERLIN\_CASA\_pipeline}} \citep{2021ascl.soft09006M}. Using a uniform image weighting, we do not find any radio emission at either epoch, with 3$\sigma$ upper limits of $195 \, \mu$Jy\,beam$^{-1}$ and $240 \, \mu$Jy\,beam$^{-1}$, respectively. However, combining the observations reduced the rms noise to $17 \, \mu$Jy\,beam$^{-1}$,\footnote{The significant reduction in the rms noise of the concatenated observation is achieved because of our ability to recover otherwise-flagged data that are flagged out during the reduction of the individual images.} enabling us to extract a significant detection of $70 \pm 8 \, \mu$Jy\,beam$^{-1}$.

\section{Results and discussion} \label{sec:Results and discussion}

The discovery of the optical counterpart \thisAT, with its rapidly fading nature and remarkably high redshift ($z = 4.859 \pm 0.002$), represents the first time an extragalactic FXT has been observed at other wavelengths. The redshift means that our optical observations sampled the emitted ultra-violet flux over the first day of its evolution in the source's rest frame.  In Table~\ref{tab:Filter wavelengths}, we present the effective wavelength centroids and widths for the filters with which we performed our observations. In the rest frame of the transient, our initial $g$-band observation with Pan-STARRS sampled $\lambda_{\rm rest} = 820_{-110}^{+120}$\,\AA, and we recovered $m_g > 23.5$ AB mag. With our subsequent redshift estimate extracted from our Gemini spectrum (see Section~\ref{sec:Discovery and follow-up - Spectra}), this non-detection is expected, since we sampled wavelengths blueward of the Lyman limit ($\lambda = 911.3$\,\AA). Even our reddest filter ($y$-band) only probed \mbox{$\lambda_{\rm rest} = 1640_{-80}^{+70}$\,\AA}, a region still well into the UV. Our dense photometric coverage, with intra-night cadence (thanks to our coordinated efforts across multiple observatories, strategically placed at different longitudes) from $T_0 + 0.054$\,d through to $T_0 + 5.537$\,d, corresponds to temporal sampling in the rest frame from $T_0 + 13.3$~minutes to $T_0 + 22.7$~hours. Our late-time Gemini observations ($T_0 + 18.42$ observer-frame days) correspond to a rest-frame phase of $T_0 + 3.14$~days.

The radio counterpart from the MeerKAT radio telescope, an unresolved point source with a flux density of $34 \pm 5 \, \mu$Jy, is also the first radio source to be associated with an extragalactic FXT. The combination of what is almost certainly non-thermal radio emission, exceptional ultra-violet luminosity, and rapid evolution indicates that \thisEP\ is most likely related to physical mechanisms that produce highly relativistic jets rather than slower thermal transients \citep[\eg,][]{2014PASA...31....8G, 2017MNRAS.466.3648A,2020SSRv..216...81A}.

\input{Tables/Filters}

\subsection{The early optical and radio fluxes} \label{Results and discussion -- Early optical and radio}

The optical discovery epoch (the cyan diamond data point in Figure~\ref{fig:AT2024eju optical lightcurve}) was obtained $\sim 520$\,s (rest frame) after the EP-WXT stopped detecting the initial X-ray emission. It then faded by $\sim 2$ magnitudes within $\approx 0.13$~rest-frame days ($m_c = 19.38 \pm 0.08$ to $m_r = 21.34 \pm 0.22$; AB mags), corresponding to a temporal index of $\approx 0.9$. The combination of proximity in time of the initial ATLAS detection and the X-ray counterpart, followed by such a rapid decay could indicate that the earliest optical emission is from the same emitting region and mechanism as the X-ray burst. Similar behaviour has been observed in some long GRBs where large field-of-view optical facilities have obtained simultaneous optical and gamma-ray detections \citep[\eg,][]{2005Natur.435..178V,2008Natur.455..183R}. It is important to distinguish between the \textit{prompt} and \textit{afterglow} emission in the optical data to create the most accurate picture of the early-time emission from this system for future modelling efforts. The afterglow component of the observed emission appears to flatten, or plateau, in the optical ($izy$) bands.

We combined our \textit{e}-MERLIN and MeerKAT detections with the published Australian Telescope Compact Array (ATCA) 5.5 and 9\,GHz $\sim 100 \, \mu$Jy detections \citep{2024GCN.35968....1L, 2024GCN.35990....1R}, and found that our observations are consistent with self-absorbed synchrotron emission. Given the spectral regime in which our data sits, we expect the radio counterpart to increase in flux density over the coming months.

\begin{figure*}
    \centering
    \subfigure{\includegraphics[width=0.48\textwidth]{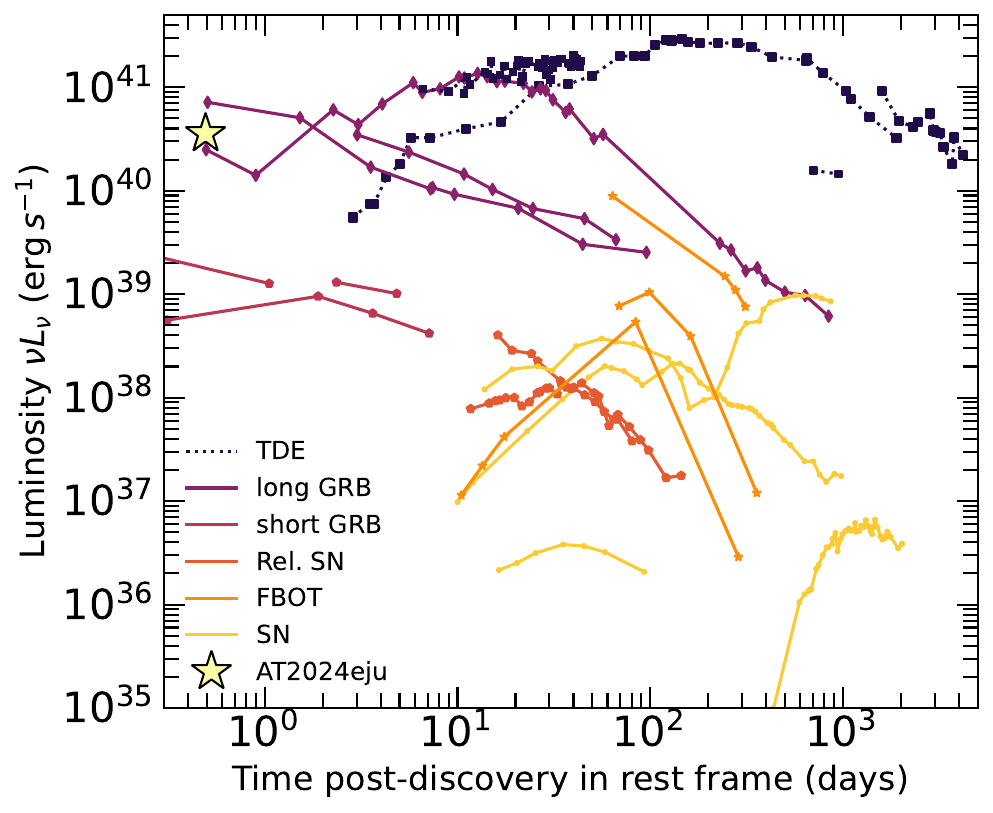}}
    \subfigure{\includegraphics[width=0.49\textwidth]{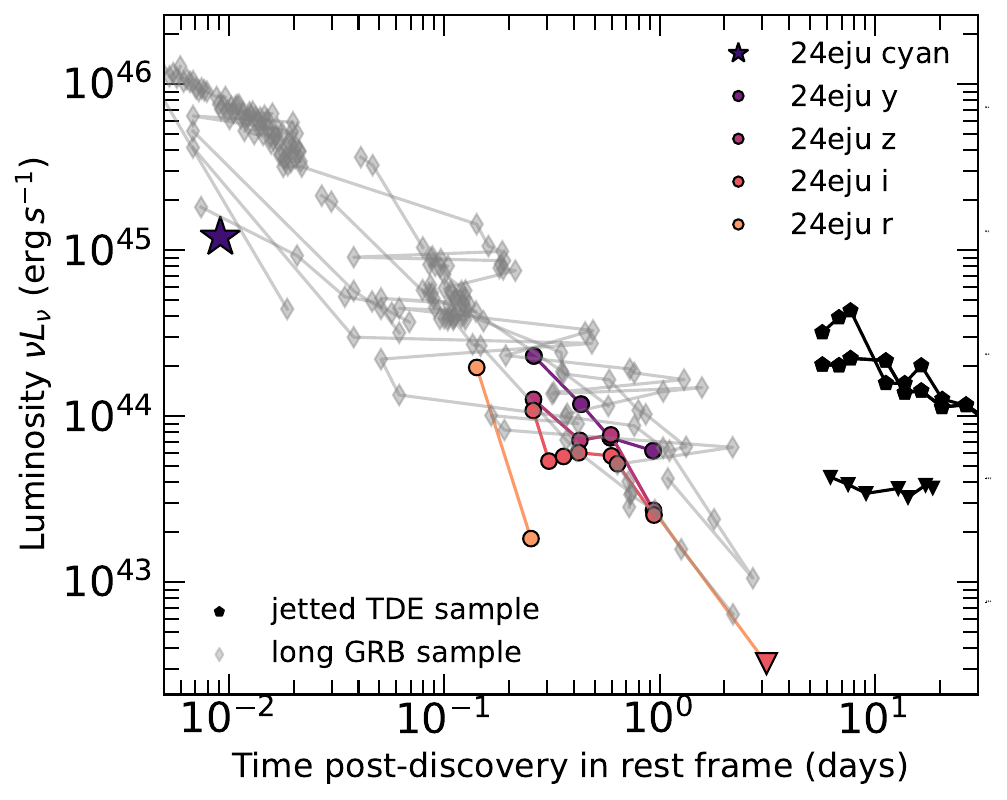}}
    \caption{
        \textit{Left panel:} Radio luminosity of different classes of extragalactic transients adapted from \citet{2020ApJ...895...49H}, including data from \citet{2023ApJ...946L..23L} and \citet{2023MNRAS.521..389R}. The yellow star indicates the radio luminosity derived from our MeerKAT detection of \thisEP. The luminosity is consistent with both GRBs and jetted TDEs.
        \textit{Right panel:} \thisAT\ optical (rest-frame UV) detections and Gemini upper limits compared to a sample of rest-frame, UV-detected, long GRB afterglows at redshifts of $z > 2$ \citep{2010ApJ...720.1513K}, the UV counterpart of the jetted TDE \ATxx{2022cmc} \citep{2024ApJ...965...39Y}, and upper limits from the candidate jetted TDE \textit{Swift} J2058.4+0516 \citep{2012ApJ...753...77C, 2022Natur.612..430A}. 
    }
    \label{fig:lum}
\end{figure*}

\subsection{What is the nature of \thisEP/\thisAT?} \label{Results and discussion - What is EP240315a?}

Despite our comprehensive follow-up campaign -- the first of its kind for an extragalactic FXT -- we cannot conclusively determine the origin of \thisEP. Our observations are consistent with two classes of extragalactic transients: gamma-ray bursts (GRBs) and jetted tidal disruption events (TDEs).

GRBs are identifiable through their highly variable prompt gamma-ray emission, followed by a smoothly evolving synchrotron afterglow, produced as a highly relativistic jet collides with the circum-burst environment. GRB\,240315C, detected by the Neil Gehrels \textit{Swift} Observatory -- Burst Alert Telescope (\textit{Swift}/BAT) and Konus Wind (KW) instruments \citep{2024GCN.35971....1D,2024GCN.35972....1S} was temporally coincident with \thisEP. The BAT signal began at $T_0 + 350$~seconds and was detected at $15 - 350$\,keV. The KW signal was detected from $T_0 + 374$~seconds at $20 - 1600$\,keV. The GRB detections lasted $\sim70$ and 47~seconds in the BAT and KW data, respectively. The IPN triangulation of GRB\,240315C places \thisEP\ just within the annulus of 26.7$^{\circ}$ width \citep{SvinkinGCN35966}. \cite{2024GCN.35972....1S} note that the KW detection had an elevated background due to increased solar flare activity. It is rare, but not completely unprecedented, to have a soft X-ray signal before the GRB itself \citep[see \eg,][]{1991Natur.350..592M, 2005ApJ...623..314P}, although in the case of \thisEP\ and GRB\,240315C the X-ray duration is significantly longer than the two previous cases. It is likely the two signals are related, but further investigation of the high energy data is required.

It is possible that \thisEP\ and GRB\,240315C fall within the class of ultra-long GRBs. Ultra-long GRBs are events whose prompt emission lasts as long as 10\,000~seconds. Like `regular' long GRBs, ultra-long GRBs have a large range of afterglow luminosities \citep[see \eg,][for multi-wavelength studies]{2014ApJ...781...13L}. As such, we cannot rule out the possibility that \thisEP\ and GRB\,240515C correspond to an ultra-long GRB event.

A short GRB (events where the prompt flash of gamma-rays is usually \emph{shorter} than $\sim$~two seconds)\footnote{This inferred duration is detector-dependent. \label{FootnoteRef}} interpretation of \thisEP\ requires a BNS merger to have occurred $\approx 1.2$\,Gyr after the Big Bang (derived from $z = 4.859$), assuming standard cosmological parameters, as adopted in Section~\ref{sec:Introduction}. Canonically dominant evolutionary channels involve an initially tight binary of massive OB stars which undergo two CCSNe \citep[\eg,][]{Tauris2017} and crucial common-envelope evolution before coalescence via gravitational inspiral. Binary evolution simulations find a range of peak delay time distributions from  $\ll 1$\,Gyr in rapid population synthesis \citep{Belczynski2018, Chruslinska2018, VignaGomez2018} to as low as 10\,Myr in the BPASS detailed stellar evolution models \citep{eldridge2019}. Additionally, the cosmic star-formation rate (and therefore the merger rate) is suppressed above $z \gtrsim 2$ \citep{MadauDickinson2014, Mapelli2018}, but many models are consistent with significant BNS merger rates at $z \approx 5$ \citep{Santoliquido2021}. Consequently, we cannot rule out a BNS origin for \thisEP\ on the grounds of stellar evolution and inspiral time alone. However, we strongly disfavour the BNS merger/short GRB scenario through the comparison of radio luminosities and timescales. A sample of radio-detected short GRB afterglows is shown in the left-hand panel of Figure~\ref{fig:lum}. Their luminosities are around two orders of magnitude lower than \thisAT, making a short GRB an unlikely origin of \thisEP/\thisAT.

Unlike short GRBs, long GRBs (events where the prompt flash of gamma-rays is usually \emph{longer} than $\sim$~two seconds),\footref{FootnoteRef} produced by collapsing massive stars,\footnote{We acknowledge the growing evidence for a population of merger-GRB events, including GRB\,211211A \citep{Rastinejad2022_GRB211211A, Troja2022_GRB211211A, Yang2022_GRB211211A, Gompertz2023_GRB211211A} and GRB\,230307A \citep{Gillanders2023_GRB230307A, Sun2023_GRB230307A, Levan2024_GRB_GRB230307A, Yang2024_GRB230307A} that lie within the long GRB $T_{90} \gtrsim 2$\,s parameter space, where $T_{90}$ is the time between the burst emitting 5 and 95~per~cent of the detected counts. However, here we are referring to the \textit{traditional} progenitor system picture, where short GRBs are produced by compact object mergers and long GRB events are produced by massive star core-collapse.} occupy a similar region of transient luminosity parameter space. In all wavebands, their afterglow component is more luminous than their short GRB analogues, likely due (at least in part) to higher kinetic energies in the jets of long GRBs compared to short GRBs \citep{2015ApJ...815..102F,2022MNRAS.511.2848A}. This is demonstrated best in the left-hand panel of Figure \ref{fig:lum}. The long GRB radio counterparts are approximately two orders of magnitude more luminous than short GRBs. As such, the long GRB radio population is far more consistent with the position of \thisAT\ in the radio luminosity parameter space. A similar conclusion can be reached regarding the position of \thisAT\ in optical luminosity parameter space. The right-hand panel of Figure \ref{fig:lum} shows a sample of optically-detected, high-redshift ($z > 2$; therefore rest-frame UV) GRB afterglows, alongside our photometric measurements of \thisAT. Their luminosities and evolution are extremely consistent.

Furthermore, the X-ray decay \citep[as reported by][]{2024GCN.35963....1L, 2024GCN.35982....1L} follows $f_{\rm X} \propto t^{-2.1}$, a decay rate which is consistent with a post-jet-break scenario \citep[which occurs when the bulk Lorentz factor of the jet is less than the inverse of the jet opening angle;][]{1999ApJ...519L..17S, 2013A&A...550L...7G, 2018ApJ...859..160W}. However, we note that the spectral slope (Photon index~$= 1.4\pm0.5$) reported by \citet{2024GCN.35951....1C} is on the hard side of afterglow spectra, but has large uncertainties \citep{2013ApJS..209...20G}.

The other main possibility for the origin of \thisEP\ is a jetted TDE, also known as a relativistic TDE. There have been two well-studied jetted TDEs, and at least two additional candidate events discovered thus far \citep{2011Natur.476..421B, 2011Sci...333..199L, 2012ApJ...753...77C, 2015MNRAS.452.4297B,2022Natur.612..430A, 2023NatAs...7...88P}. Whilst not all jetted TDEs have optical counterparts, they all have luminous and highly variable X-ray counterparts. In the right-hand panel of Figure~\ref{fig:lum}, we present the values and limits of rest-frame optical/UV jetted TDE observations (in luminosity space), alongside the GRB rest-frame UV detections. Furthermore, all jetted TDEs so far have luminous, long-lasting radio counterparts, consistent with highly relativistic jets \citep[\eg,][]{2011Natur.476..425Z, 2023MNRAS.521..389R}.

At a redshift of $z = 4.859$, the isotropic X-ray luminosity of \thisEP\ \citep[from the average unabsorbed $0.5 - 4.0$\,keV flux of \mbox{$5.3_{-0.7}^{+1.0} \times 10^{-10}$\,\ergscm}, as reported by][]{Zhang2024_GCN35931} is \mbox{$L_{\rm X} \simeq 1.3 \pm 0.2 \times 10^{50}$\,erg\,s$^{-1}$}, over the rest-frame $3 - 23$\,keV band. This sits at the top end of the luminosity range for the X-ray flares associated with \textit{Swift}~J1644. The X-ray decay and the photon index measurements are also similar \citep{2011Natur.476..421B, 2024GCN.35982....1L}. The first optical data point of \ATxx{2022cmc} from \citet{2022Natur.612..430A} was acquired one day post-burst (rest frame), which is later than the detections of \thisAT\ we report here. However, extrapolation of the \ATxx{2022cmc} detections shows that they are consistent with the results we report here for \thisAT.

At radio frequencies, both \textit{Swift}~J1644 and \ATxx{2022cmc} have reported luminous, slowly evolving counterparts, as shown in the left-hand panel of Figure~\ref{fig:lum}. The radio emission comes from external shocks between the jet and the circum-nuclear environment. The radio detection of \thisAT, whilst made earlier than for the other jetted TDEs presented, occupies the same luminosity parameter space.

\thisEP\ has characteristics that would allow it to be classified as either a GRB or a jetted TDE. Figure~\ref{fig:lum} illustrates where our radio and optical discoveries sit compared to other extragalactic transients that have been detected in both the radio and optical bands. Currently, it is not possible to differentiate between the TDE or GRB scenarios. Radio observations of both GRBs and jetted TDEs have found optically thick counterparts at early times \citep{2023NatAs...7..986B, 2023MNRAS.521..389R}, consistent with our findings here. 

In the rest-frame UV, all of our detections are consistent with the low-luminosity end of the GRB afterglow distribution. There are no detections of jetted TDEs at such early times. The earliest UV detection of a jetted TDE recorded is $\sim$~a few days post-burst \citep[rest frame;][]{2024ApJ...965...39Y}, later than our final optical detection. Furthermore, only one jetted TDE has been detected in the UV. The other event only has upper limits \citep{2012ApJ...753...77C}, an order of magnitude below the detections. This range demonstrates the large possible range of UV parameter space associated with TDEs that is still to be explored, making it very hard to estimate the early UV properties of the jetted TDE family.

We rule out the possibility of \thisAT\ being an FBOT-like transient based on the mismatch between the early evolution of this FXT and that of \ATxx{2018cow}, the prototypical FBOT transient. While the peak bolometric luminosity of \ATxx{2018cow} roughly matches the early follow-up observations of \thisAT\ \citep[\mbox{$L_{\rm bol} \sim 10^{44}$\,erg\,s$^{-1}$;}][]{Prentice2018}, the rise time is much slower; $t_{\rm rise} \simeq 3$~rest-frame days, versus \mbox{$\simeq 0.3$~rest-frame days} for \thisAT. Additionally, the early radio lightcurve of \ATxx{2018cow} demonstrates a slow rise to maximum radio luminosity of $\sim 100$~rest-frame days \citep[see \eg,][]{Ho2019}, which does not match the early, very luminous radio detection for \thisEP. This FXT event evolves on a much more rapid timescale than \ATxx{2018cow}, both in the optical and radio, leading us to rule out FBOTs as an explanation for this transient.

\subsection{Archival search for other orphan fast-evolving optical transients}

Fast-fading transients are commonly found by ATLAS and the Zwicky Transient Facility 
\citep[ZTF;][]{2019PASP..131a8002B}, some of which may be extragalactic counterparts to GRBs or FXTs  \citep{2017ApJ...850..149S, 2021ApJ...918...63A}. However, the main issue with identifying such transients is the foreground contamination rate of fast cataclysmic variables (CVs), which also often have no host star in Pan-STARRS or Legacy Survey images. In the ATLAS database, there have been $\sim 400$ objects flagged as high-significance transients with no catalogued host \citep{Smith2020_ATLAS}.

We manually checked all of these objects and found that most had decay rates that were too slow to match \thisAT\ (likely supernovae or relatively common CVs), or were characteristic of stellar variability with low signal-to-noise ratios. We found 34 genuine orphans detected on only one night, with signs of rapid fading (evidenced by a non-detection in quick succession to the sole epoch of detection). However, the constraints on their rate of fading still do not allow them to be confidently separated from Galactic CVs. Only by combining with external triggers, such as the Einstein Probe, will we be able to build a better understanding of the optical counterparts to extragalactic FXTs.

\section{Summary and conclusions} \label{sec:Summary and conclusions}

In this Letter we presented the discovery of the optical and radio counterparts to the Einstein Probe FXT, \thisEP.

The optical counterpart, \thisAT, was detected as a host-less transient by ATLAS during its routine all-sky survey operations just 0.054~days (1.28~hours) after the X-ray signal recorded by the Einstein Probe. We recorded a non-detection 1.7~days prior to discovery, and constrained its rapid fade, with it decaying by $\sim 2$ magnitudes in 19~hours post-discovery (see Figure~\ref{fig:AT2024eju detection images}).

The radio counterpart to \thisEP\ was discovered by the MeerKAT radio telescope 2.86~days after the X-ray signal was recorded, and has been shown to originate from optically thick synchrotron emission by follow-up complementary \textit{e}-MERLIN observations.

Our measured redshift ($z = 4.859 \pm 0.002$) rules out some of the models proposed for FXTs, including supernova shock breakouts, binary neutron star mergers and tidal disruption of white dwarfs. The inferred high luminosity of the X-ray, optical and radio emission implies that this is a relativistic event, and we propose two possible scenarios: a long GRB or a jetted TDE. 

To differentiate between the GRB and TDE scenarios, continued monitoring of \thisEP\ is needed in both the radio and optical bands. The evolutionary timescale of the afterglow is a clear differentiating feature of GRBs from TDEs, as GRBs evolve much more rapidly -- usually decaying on a timescale of $\sim {\rm days} - {\rm weeks}$. If we consider the optical temporal behaviour of \ATxx{2022cmc} to be characteristic of all jetted TDEs, then the TDE lightcurve should plateau. Our Gemini upper limits, obtained at $T_0 +18.42$~days post-discovery, are deep enough to rule out an \ATxx{2022cmc}-like lightcurve and luminosity over the same temporal range. The Gemini upper limits favour a GRB-like lightcurve, but are still consistent with the upper limits obtained for \textit{Swift} J1644 \citep{2011Sci...333..203B}.

In the radio band, we predict that differentiating between the TDE and GRB scenarios will take longer, at least $50 - 100$~rest-frame days. The decay rate inferred from the Gemini upper limits seems to prefer the GRB scenario (see Figure~\ref{fig:lum}). However, given the lack of knowledge regarding UV counterparts to jetted TDEs, we cannot confidently rule out the TDE scenario without further observations.

Whilst the majority of long GRBs have been observed to decay in the radio, there are exceptions such as GRB\,030329 \citep{2003Natur.426..154B, 2008A&A...480...35V}, where they continued to rise for weeks post-burst. Fortunately, given the high luminosity of the radio counterpart, even at a redshift of $z = 4.859 \pm 0.002$, it will be possible to track the radio emission for months -- years, allowing us to confidently classify this transient with future observations.

While it is difficult to quantify the future rates of such events, the discovery of \thisEP\ so soon after the launch of the EP ($\sim 2$~months) indicates that such events are probably not intrinsically rare. The soft X-ray regime that EP is optimised to explore is ideal for searching for high-redshift events that emit high-energy radiation, as redshifting will shift the peak of this emission from $\gamma$- to X-rays, making them more detectable to the EP, and other X-ray observatories. The nature of this FXT indicates the Einstein Probe will uncover a range of high-energy transient phenomena in both the low- and high-redshift Universe.

\input{Tables/Photometry}

\section*{Acknowledgments}

We thank the anonymous referee for useful comments that helped improve the final version of the manuscript.

We thank Andrew~J.~Bunker and Alex~J.~Cameron for useful discussions regarding the interpretation of the redshift of \thisEP. 

We thank L.~Piro, G.~Bruni, A.~Linesh Thakur and G.~Gianfagna for useful discussion surrounding the initial reduction of the e-MERLIN data, which was jointly awarded under two distinct program IDs (DD17003 \& DD17004; PIs: G.~Bruni, L.~Rhodes, respectively).

SJS, SS, KWS and DRY acknowledge funding from STFC Grants ST/Y001605/1, ST/X001253/1, ST/X006506/1 and ST/T000198/1. 
SJS acknowledges a Royal Society Research Professorship. 
HFS is supported by the Eric and Wendy Schmidt AI in Science Fellowship.
RF acknowledges support from STFC, the ERC and the Hintze charitable foundation.
PGJ has received funding from the European Research Council (ERC) under the European Union’s Horizon 2020 research and innovation programme (Grant agreement No.~101095973).
MN is supported by the European Research Council (ERC) under the European Union's Horizon 2020 research and innovation programme (grant agreement No.~948381) and by UK Space Agency Grant No.~ST/Y000692/1. 
FC acknowledges support from the Royal Society through the Newton International Fellowship programme (NIF/R1/211296).
AJC acknowledges support from the Hintze Family Charitable Foundation.
TWC acknowledges the Yushan Young Fellow Program by the Ministry of Education, Taiwan for the financial support. 
SY acknowledges the funding from the National Natural Science Foundation of China under Grant No. 12303046.

ATLAS is primarily funded through NASA grants NN12AR55G, 80NSSC18K0284, and 80NSSC18K1575. The ATLAS science products are provided by the University of Hawaii, Queen’s University Belfast, STScI, SAAO and Millennium Institute of Astrophysics in Chile. 

Pan-STARRS is primarily funded to search for near-earth asteroids through NASA grants NNX08AR22G and NNX14AM74G. The Pan-STARRS science products for transient follow-up are made possible through the contributions of the University of Hawaii Institute for Astronomy and Queen's University Belfast.

This publication has made use of data collected at Lulin Observatory, partly supported by MoST grant 108-2112-M-008-001. We thank Lulin staff H.-Y.~Hsiao, W.-J.~Hou, C.-S.~Lin, H.-C.~Lin, and J.-K.~Guo for observations and data management, and C.-Y.~Cheng for fringing pattern exam. 

The Liverpool Telescope is operated on the island of La Palma by Liverpool John Moores University in the Spanish Observatorio del Roque de los Muchachos of the Instituto de Astrofisica de Canarias with financial support from the UK Science and Technology Facilities Council.

Based on observations obtained at the international Gemini Observatory (under program IDs GN-2024A-Q-221 and GN-2024A-Q-128), a program of NSF NOIRLab, which is managed by the Association of Universities for Research in Astronomy (AURA) under a cooperative agreement with the U.S. National Science Foundation on behalf of the Gemini Observatory partnership: the U.S. National Science Foundation (United States), National Research Council (Canada), Agencia Nacional de Investigaci\'{o}n y Desarrollo (Chile), Ministerio de Ciencia, Tecnolog\'{i}a e Innovaci\'{o}n (Argentina), Minist\'{e}rio da Ci\^{e}ncia, Tecnologia, Inova\c{c}\~{o}es e Comunica\c{c}\~{o}es (Brazil), and Korea Astronomy and Space Science Institute (Republic of Korea).
This work was enabled by observations made from the Gemini North telescope, located within the Maunakea Science Reserve and adjacent to the summit of Maunakea. We are grateful for the privilege of observing the Universe from a place that is unique in both its astronomical quality and its cultural significance.

The MeerKAT telescope is operated by the South African Radio Astronomy Observatory, which is a facility of the National Research Foundation, an agency of the Department of Science and Innovation. This work has made use of the ``MPIfR S-band receiver system'' designed, constructed and maintained by funding of the MPI f{\"u}r Radioastronomy and the Max-Planck-Society.

e-MERLIN is a National Facility operated by the University of Manchester at Jodrell Bank Observatory on behalf of STFC.

\facilities{
    \begin{itemize}
        \item   ATLAS
        \item   Pan-STARRS
        \item   Gemini
        \item   Liverpool:2m
        \item   MeerKAT
        \item   \textit{e}-MERLIN
        \item   LO:1m
    \end{itemize}
}

\software{
    \begin{itemize}
        \item   \texttt{Astropy}      \citep{2013A&A...558A..33A,2018AJ....156..123A,2022ApJ...935..167A}
        \item   \texttt{Matplotlib}   \citep{Hunter2007_matplotlib}
        \item   \texttt{NumPy}        \citep{Harris2020}
        \item   \texttt{pandas}       \citep{mckinney-proc-scipy-2010,reback2020pandas}
        \item   \texttt{DRAGONS}      \citep{Labrie2023_DRAGONS, DRAGONS_zenodo}
        \item   \texttt{CASA}         \citep{CASA_team_2022}
        \item   \texttt{OxKAT}        \citep{oxkat}
    \end{itemize}
}

\bibliography{References}
\bibliographystyle{aasjournal}

\end{document}

%% file: Commands.tex
\newcommand{\eg}{\mbox{e.g.}}

\newcommand{\SNxx}[1]{\mbox{SN\,#1}}
\newcommand{\ATxx}[1]{\mbox{AT\,#1}}

\newcommand{\EPxx}[1]{\mbox{EP\,#1}}

\newcommand{\thisEP}{\EPxx{240315a}}
\newcommand{\thisAT}{\ATxx{2024eju}}

\newcommand{\I}{{\sc i}}

\newcommand{\IV}{{\sc iv}}
\newcommand{\V}{{\sc v}}

\newcommand{\grizy}{\ensuremath{grizy}}

\newcommand{\izy}{\ensuremath{izy}}

\newcommand{\ergscm}{erg\,s$^{-1}$\,cm$^{-2}$}

%% file: Packages.tex
\usepackage{multirow}

\usepackage{booktabs}
\usepackage{subfigure}

\usepackage{xcolor}

%% file: Tables/Filters.tex

\begin{table*}
    \renewcommand*{\arraystretch}{1.1}
    \centering
    \caption{
        Wavelength coverage for the different filters/bands of our observations (both optical and radio). We present the wavelengths/frequencies in both the observer and rest frames, to emphasise the short wavelengths/high frequencies being sampled in the rest frame of the transient. The ATLAS $co$-band wavelength information is taken from \cite{Tonry2018_ATLAS}, while the Pan-STARRS \grizy-band wavelength information is extracted from \cite{Tonry2012_PS_photometry}.
    }
    \begin{tabular}{ccccc}
        \toprule
        \multicolumn{5}{c}{Optical}    \\
        \midrule

        Filter      &\multicolumn{2}{c}{Wavelength}                       &\multicolumn{2}{c}{Wavelength}            \\
                    &\multicolumn{2}{c}{(\AA, observer frame)}            &\multicolumn{2}{c}{(\AA, rest frame)}     \\
        
        \midrule

        $c$ 	 &\multicolumn{2}{c}{$5330_{- 1100}^{+ 1180}$}       &\multicolumn{2}{c}{$910_{- 190}^{+ 200}$}      \\
        $o$ 	 &\multicolumn{2}{c}{$6780_{- 1170}^{+ 1410}$}       &\multicolumn{2}{c}{$1160_{- 200}^{+ 240}$}     \\

        $g$ 	 &\multicolumn{2}{c}{$4810_{- 670}^{+ 700}$}         &\multicolumn{2}{c}{$820_{- 110}^{+ 120}$}      \\
        $r$ 	 &\multicolumn{2}{c}{$6170_{- 670}^{+ 720}$}         &\multicolumn{2}{c}{$1050_{- 110}^{+ 120}$}     \\
        $i$ 	 &\multicolumn{2}{c}{$7520_{- 620}^{+ 670}$}         &\multicolumn{2}{c}{$1280_{- 100}^{+ 110}$}     \\
        $z$ 	 &\multicolumn{2}{c}{$8660_{- 480}^{+ 560}$}         &\multicolumn{2}{c}{$1480_{- 80}^{+ 100}$}      \\
        $y$ 	 &\multicolumn{2}{c}{$9620_{- 440}^{+ 390}$}         &\multicolumn{2}{c}{$1640_{- 80}^{+ 70}$}       \\

        \midrule
        \multicolumn{5}{c}{Radio}    \\
        \midrule

        Filter      &Central frequency       &Bandwidth   &Central frequency    & Bandwidth             \\
                    &(GHz, observer frame)   &            &(GHz, rest frame)    &                       \\
        \midrule
        S3          &$3.06$   &$0.88$   &$17.93$     &$5.16$          \\
        C           &$5.01$   &$0.51$   &$29.35$     &$2.99$          \\

        \bottomrule
    \end{tabular}
    \label{tab:Filter wavelengths}
\end{table*}

%% file: Tables/Photometry.tex

\begin{table*}
    \renewcommand*{\arraystretch}{1.1}
    \centering
    \caption{
        Optical and radio photometry of \thisAT. Magnitudes have not corrected for the expected foreground extinction of \mbox{$E (B - V) = 0.042$} \citep{Schlafly2011}. The errors for the optical photometry are quoted to $1 \sigma$, while upper limits are quoted to $2 \sigma$ significance. The radio upper limits are quoted to $3 \sigma$. The final radio column contains the result of concatenating the two \textit{e}-MERLIN non-detections, resulting in a radio detection. 
    }
    \begin{tabular}{cccccc}
        \toprule
        \multicolumn{6}{c}{Optical}       \\
        \midrule
        
        $T_{\rm mid} - T_0$       &MJD       &Telescope       &Total exposure         &Filter        &Apparent magnitude     \\
        (days)          &          &                &time (s)               &              &(AB mag)               \\
        \midrule
        
        $-3.950$ 	 &60380.891 	 &ATLAS	       &120 	     &$c$ 	 &$> 21.6$        	 \\ 
        $-1.696$ 	 &60383.145 	 &ATLAS	       &120 	     &$o$ 	 &$> 20.2$        	 \\ 
        $+0.054$ 	 &60384.894 	 &ATLAS	       &120 	     &$c$ 	 &$19.38 \pm 0.08$ 	 \\ 
        $+0.832$ 	 &60385.673 	 &SLT 	       &1800     	 &$r$ 	 &$21.34 \pm 0.22$ 	 \\ 
        $+1.007$ 	 &60385.848 	 &LT 	       &180 	     &$g$ 	 &$> 20.0$      	 \\ 
        $+1.010$ 	 &60385.851 	 &LT 	       &180 	     &$r$ 	 &$> 20.1$      	 \\ 
        $+1.012$ 	 &60385.853 	 &LT 	       &180 	     &$i$ 	 &$> 20.3$      	 \\ 
        $+1.304$ 	 &60386.144 	 &ATLAS	       &120 	     &$o$ 	 &$> 21.1$        	 \\ 
        $+1.483$ 	 &60386.324 	 &PS 	       &300 	     &$g$ 	 &$> 23.5$      	 \\ 
        $+1.486$ 	 &60386.327 	 &PS 	       &300 	     &$r$ 	 &$23.92 \pm 0.49$ 	 \\ 
        $+1.522$ 	 &60386.363 	 &PS 	       &700 	     &$i$ 	 &$21.99 \pm 0.06$ 	 \\ 
        $+1.527$ 	 &60386.368 	 &PS 	       &700 	     &$z$ 	 &$21.82 \pm 0.07$ 	 \\ 
        $+1.531$ 	 &60386.372 	 &PS 	       &700 	     &$y$ 	 &$21.17 \pm 0.10$ 	 \\ 
        $+1.801$ 	 &60386.642 	 &LOT 	       &3000     	 &$i$ 	 &$22.75 \pm 0.23$ 	 \\ 
        $+2.105$ 	 &60386.946 	 &LT 	       &2400     	 &$i$ 	 &$22.68 \pm 0.37$ 	 \\ 
        $+2.135$ 	 &60386.976 	 &LT 	       &2400     	 &$z$ 	 &$> 22.0$         	 \\ 
        $+2.477$ 	 &60387.318 	 &PS 	       &2000     	 &$i$ 	 &$22.62 \pm 0.06$ 	 \\ 
        $+2.505$ 	 &60387.346 	 &PS 	       &2000     	 &$z$ 	 &$22.44 \pm 0.08$ 	 \\ 
        $+2.535$ 	 &60387.375 	 &PS 	       &1600     	 &$y$ 	 &$21.89 \pm 0.15$ 	 \\ 
        $+3.466$ 	 &60388.307 	 &PS 	       &1600     	 &$y$ 	 &$22.40 \pm 0.23$ 	 \\ 
        $+3.491$ 	 &60388.331 	 &PS 	       &2000     	 &$z$ 	 &$22.36 \pm 0.08$ 	 \\ 
        $+3.515$ 	 &60388.356 	 &PS 	       &2000     	 &$i$ 	 &$22.67 \pm 0.07$ 	 \\ 
        $+3.732$ 	 &60388.573 	 &SLT 	       &8700     	 &$i$ 	 &$> 21.8$        	 \\ 
        $+3.747$ 	 &60388.588 	 &LOT 	       &9000     	 &$i$ 	 &$22.79 \pm 0.36$ 	 \\ 
        $+4.717$ 	 &60389.558 	 &LOT 	       &6000     	 &$i$ 	 &$> 22.1$        	 \\ 
        $+5.477$ 	 &60390.317 	 &PS 	       &2400     	 &$y$ 	 &$22.59 \pm 0.29$ 	 \\ 
        $+5.507$ 	 &60390.347 	 &PS 	       &2400     	 &$z$ 	 &$23.49 \pm 0.32$ 	 \\ 
        $+5.537$ 	 &60390.378 	 &PS 	       &2400     	 &$i$ 	 &$23.56 \pm 0.31$ 	 \\ 
        $+18.42$ 	 &60403.257 	 &Gemini       &910 	     &$i$ 	 &$> 25.8$       	 \\ 
        $+18.44$ 	 &60403.276 	 &Gemini       &1170     	 &$z$ 	 &$> 25.8$      	 \\ 
                
        \midrule
        \multicolumn{6}{c}{Radio}       \\
        \midrule

        $T_{\rm mid} - T_0$       &MJD       &Telescope            &Central              &Flux density                 &Concatenated flux                   \\
        (days)                      &          &                     &frequency (GHz)      &($\mu$Jy\,beam$^{-1}$)       &density ($\mu$Jy\,beam$^{-1}$)      \\
        \midrule

        $+2.86$            &60387.703      &MeerKAT                &3.06      &$34 \pm 5$     &$-$                              \\
        
        $+5.57$            &60390.407      &\textit{e}-MERLIN      &5.01      &$< 195$        &\multirow{2}{*}{$70 \pm 8$}      \\
        $+12.54$           &60397.375      &\textit{e}-MERLIN      &5.01      &$< 240$        &                                 \\        

        \bottomrule
    \end{tabular}
    \label{tab:Photometry}
\end{table*}